\documentclass[secnumarabic, amssymb, aps]{revtex4-2}
\usepackage{epstopdf}
\usepackage[utf8]{inputenc}
\usepackage{amsmath,amssymb,amsthm,amsfonts,mathrsfs,bm,verbatim,hyperref}
\usepackage{graphicx,subfigure}
\usepackage[section]{placeins}
\usepackage{ulem}
\newcommand{\bea}{\begin{eqnarray}}
\newcommand{\eea}{\end{eqnarray}}

\renewcommand\thesection{\arabic{section}}

\def\bea{\begin{aligned}}
\def\ena{\end{aligned}}

\allowdisplaybreaks[4]

\UseRawInputEncoding

\begin{document}

\title{Shadows and photon rings of a quantum black hole}

\author{Jing-Peng Ye $^{1}$}
\author{Zhi-Qing He$^{1}$}
\author{Ai-Xu Zhou$^{2}$}
\author{Zi-Yang Huang$^{2}$}
\author{Jia-Hui Huang$^{2,3}$}
\email{huangjh@m.scnu.edu.cn}

\affiliation{$^1$ School of Information and Optoelectronic Science and Engineering, South China Normal University, Guangzhou 510631, China\\
$^2$ Key Laboratory of Atomic and Subatomic Structure and Quantum Control (Ministry of Education), Guangdong Basic Research Center of Excellence for Structure and Fundamental Interactions of Matter, School of Physics, South China Normal University, Guangzhou 510006, China\\
$^3$ Guangdong Provincial Key Laboratory of Quantum Engineering and Quantum Materials, Guangdong-Hong Kong Joint Laboratory of Quantum Matter, South China Normal University, Guangzhou 510006, China}

\begin{abstract}
Recently, a black hole model in loop quantum gravity has been proposed by Lewandowski, Ma, Yang and Zhang (Phys. Rev. Lett. \textbf{130}, 101501 (2023)). The metric tensor of the quantum black hole (QBH) is a suitably modified Schwarzschild one. In this paper, we calculate the radius of the circular null geodesic (light ring) and obtain the linear approximation of it with respect to the quantum correction parameter $\alpha$: $r_{l} \simeq 3 M - \frac{\alpha}{9 M}$. We then assume the QBH is backlit by a large, distant plane of uniform, isotropic emission and calculate the radius of the black hole shadow and its linear approximation: $r_{s} = 3 \sqrt{3} M - \frac{\alpha}{6 \left(\sqrt{3} M\right)}$. We also consider the photon ring structures in the shadow when the impact parameter $b$ of the photon approaches to a critical impact parameter $b_{\textrm{c}}$, and obtain a formula for estimating the deflection angle, which is $\varphi_{\textrm{def}} = - \frac{\sqrt{2}}{\omega r_{l}^2}\log{\left(b - b_c\right) + \widetilde{C}(b)}$. We also numerically plot the images of shadows and photon rings of the QBH in three different illumination models and compare them with that of a Schwarzschild black hole. It is found that we could distinguish the quantum black hole with a Schwarzschild black hole via the shadow images in certain illumination models.
\end{abstract}

\maketitle

\section{Introduction}\label{section1}
Although the gravitational force can now be well described by classical general relativity, there exist some problems for general relativity. Singularity is one of the most fundamental ones. Penrose proposed the first version of the singularity theorem \cite{SingPenrose}, and a more general one was proved by Hawking and Penrose \cite{SingHawking}, which states that the existence of spacetime singularity is inevitable under certain common physical conditions.
How should one treat spacetime singularities? We may expect that a quantum theory of gravity could cure spacetime singularities. One of the candidate theories of quantum gravity is loop quantum gravity (LQG), which is a background-independent and nonperturbative scheme \cite{Rovelli,Thiemann, Thiemann:2002nj,Ashtekar:2004eh,Han:2005km,Giesel:2012ws,Rovelli:2011eq,Perez:2012wv}. The cosmological big-bang singularity was resolved theoretically and numerically in the context of loop quantum cosmology(LQC) \cite{ Bojowald:2001xe,Ashtekar:2003hd,Ashtekar:2006rx,Ashtekar:2006uz,Ashtekar:2006wn}. For the singularity of a Schwarzschild black hole (BH), some attempts aimed to solve it by quantizing the interior of the BH with the techniques developed in LQG \cite{Zhang:2021wex,Zhang:2020qxw,Bodendorfer:2019cyv,Ashtekar:2018cay,Ashtekar:2018lag,Corichi:2015xia,Modesto:2008im,Boehmer:2007ket,Ashtekar:2005qt}. In addition, the LQG corrections to BH formation or gravitational collapse in different models have also been studied \cite{Giesel:2021dug,Husain:2021ojz,Munch:2021oqn,Munch:2020czs,BenAchour:2020gon,BenAchour:2020bdt,Kelly:2020lec,Bojowald:2009ih,Bojowald:2005qw,Yang:2022btw,Lewandowski:2022zce}.

Recently, the quantum effects of LQG have been taken into account in studying the gravitational collapse of spherically symmetric dust matter \cite{Kelly:2020lec,Yang:2022btw,Lewandowski:2022zce}. The dust collapse was studied in the context of LQC model. The collapse stopped when the energy density of the dust reached the Plank scale, and the collapsing dust then turned to bounce. By matching metrics at the boundary between the interior and exterior of the collapsing dust, the LQG corrected exterior vacuum metric was derived. In the collapsing stage, a quantum-corrected Schwarzschild metric appeared in the exterior spacetime, while, in the late bouncing stage, something like a white hole appeared \cite{Yang:2022btw}. 

Significant progresses have also been made for observation of black holes via radio waves in recent years. The first image of the supermassive black hole $\text{M}87^*$ was provided by Event Horizon Telescope in 2019 \cite{M87-EHT-1,M87-EHT-4}. In 2022, the image of the supermassive black hole Sagittarius $A^*$ in our Galactic Center was also released \cite{SgrA-EHT-1,SgrA-EHT-3}. Two important concepts  related to these images are the light ring and shadow of a BH. A light ring is a circular photon orbit around a BH  with a radius larger than the event horizon \cite{Cunha:2020azh,Guo:2020qwk,Wei:2020rbh,Ghosh:2021txu}. A shadow is a dark area casted by a black hole for an observer at asymptotic infinity. The radius of a BH shadow is closely related to the radius of the light ring \cite{Cunha:2017eoe,cunha2015,Gralla:2019xty,Cvetic:2016bxi,Lu:2019zxb,Ma:2019ybz,Yang:2019zcn,Wielgus:2021peu,ISCOaS4DEGBBH-GMY-LPC,Gauss,abb579,	kumar2020,shaikh,weiliu,zhongchen,lichen,Shaikh2019,Chowdhuri:2020ipb,AbhishekChowdhuri:2023ekr,Ghosh:2022mka,Anacleto:2021qoe,Tsukamoto:2014tja,Tsukamoto:2017fxq,
Konoplya:2019xmn,Younsi:2016azx,Bambi:2019tjh,Vagnozzi:2019apd,Allahyari:2019jqz}. Usually, the shadow is not totally dark and there are some tiny bright structures in it, which is dubbed photon rings \cite{Gralla:2019xty}.   
Obviously, the geometric information of a BH can be decoded from the structures of the bright photon rings and shadow of the BH.

In this paper, we will investigate the shadow and the structure of the bright photon rings of the quantum corrected black hole (QBH) proposed in \cite{
Lewandowski:2022zce,Yang:2022btw}.
In section II, we calculate the radius of the light ring and the critical impact parameter for the QBH case, discuss their deviations from the Schwarzschild BH case and provide the leading term of the deviation with respect to the quantum correction parameter $\alpha$. 
In section III,  we consider a model where the QBH is backlit by a large, distant screen of uniform, isotropic emission \cite{Gralla:2019xty}, obtain the radius of shadow under the background light source, and discuss the relation of deflection angle and the impact parameter. 
In section IV,  we plot the shadows and photon rings of the QBH under several illumination conditions, and compare them with that of a Schwarzschild black hole (SBH). We summarize our work in section V.

\section{Radius of light ring}\label{section2}
In this section, we discuss the radius of the unstable light ring of the QBH spacetime proposed in \cite{Yang:2022btw,Zhang:2023okw} with the metric described by
\begin{eqnarray}\label{LQGBH2}
        ds_{}^2 = -f(r) dt^2 + f(r)^{-1} dr^2 + r^2 (d\theta^2+\sin^2\theta d\varphi^2),~ f(r) = 1 - \frac{2 M}{r} + \frac{\alpha  M^2}{r^4}.
\end{eqnarray}
Note that discussions on other quantum corrected black holes can be found in \cite{Jusufi:2022uhk,Battista:2023iyu,Afrin:2022ztr,Vagnozzi:2022moj}. Here we use the natural units, $ c = G = \hbar= 1 $.  $\alpha=16\sqrt{3}\pi \gamma^3$ is the quantum correction parameter. $ \gamma \approx 0.2375 $ is the Barbero-Immirzi parameter \cite{Meissner:2004ju,Domagala:2004jt}. There is a lower bound $ M_{\textrm{min}} $ for the mass parameter $M$ of the above BH in order that the horizons exist. In the natural units, the value of this lower bound is  
 \begin{equation}\label{massLowerBoundHorizon}
         M_{\textrm{min}} = \frac{16 \gamma \sqrt{\pi \gamma}}{3\sqrt[4]{3}} \approx 0.8314
 \end{equation}
 which is a little smaller than the Plank mass $M_P=1$.
In this paper we will focus on $ M > M_{\textrm{min}} $, there are two different horizons, one inner and one outer horizon, which are denoted by $ r_{-} $ and $ r_{+} $ respectively. 

Due to the spherical symmetry of the QBH, a planar null geodesic motion is equivalent to an equatorial null geodesic motion. The null trajectory equation in the equatorial plane is given by \cite{Chen:2022ewe,MTofBH} 
\begin{equation}\label{governingEquationOfPhoton}
    \frac{1}{r^4} \left(\frac{dr}{d\varphi}\right)^{2} = \frac{E^{2}}{L^{2}} - U_{\textrm{MS}}(r)=\left(\frac{1}{b}\right)^{2} - U_{\textrm{MS}}(r),
\end{equation}
where $E$ and $L$ are conserved energy and orbital angular momentum respectively. $b=\frac{L}{E}$ is the impact parameter. $ U_{\textrm{MS}}(r) $ is the effective potential
\begin{equation}\label{effectivePotentialOfPhoton}
    U_{\textrm{MS}}(r) = \frac{1}{r^2} f(r) = \frac{1}{r^{2}}\left(1 - \frac{2 M}{r} + \frac{\alpha M^{2}}{r^{4}}\right).
\end{equation}
The possible values of the radii of the light rings are  determined by the extremal values of the effective potential $U_{\textrm{MS}}$, i.e. the roots of the following equation
 \begin{equation}\label{eq7}
          \frac{dU_{\textrm{MS}}}{dr} = 0
 \end{equation}
Since the above equation is a quartic polynomial equation, all the four roots can be given in analytic forms. It is found that when $M> \frac{2^{4}}{3^{3}} \sqrt{\alpha} \approx 0.64$ ,two real roots exist for the equation. So we have two real roots and two complex roots when the mass $M$ is greater than $ M_{\textrm{min}}\approx 0.8314 $. The smaller real root is located inside the outer horizon $ r_{+} $, so the radius of the light ring is 
 \begin{eqnarray}
r_{l} = \frac{1}{4}\left(3 M + 2 \xi + \sqrt{\frac{\left(3 M - \xi\right)\left(3 M + 2 \xi\right)^{2}}{\xi}}\right),
 \end{eqnarray}\label{rLR}
where
\begin{equation}\label{eq17}
    \xi=\sqrt{\frac{9 M^{2}}{4}+\sigma},
\end{equation}
and 
\begin{eqnarray}
\sigma = \frac{4 M^2 \alpha}{\sqrt[3]{\frac{27 M^4 \alpha}{2} - \frac{1}{2} \sqrt{M^4 \alpha^2(729 M^4 - 256 M^2 \alpha)}}}
+ \sqrt[3]{\frac{27 M^4 \alpha}{2} - \frac{1}{2} \sqrt{M^4 \alpha^2 (729 M^4 - 256 M^2 \alpha)}}.
\end{eqnarray}

\begin{figure}[tbph]
\centering
\includegraphics[width=8.5cm]{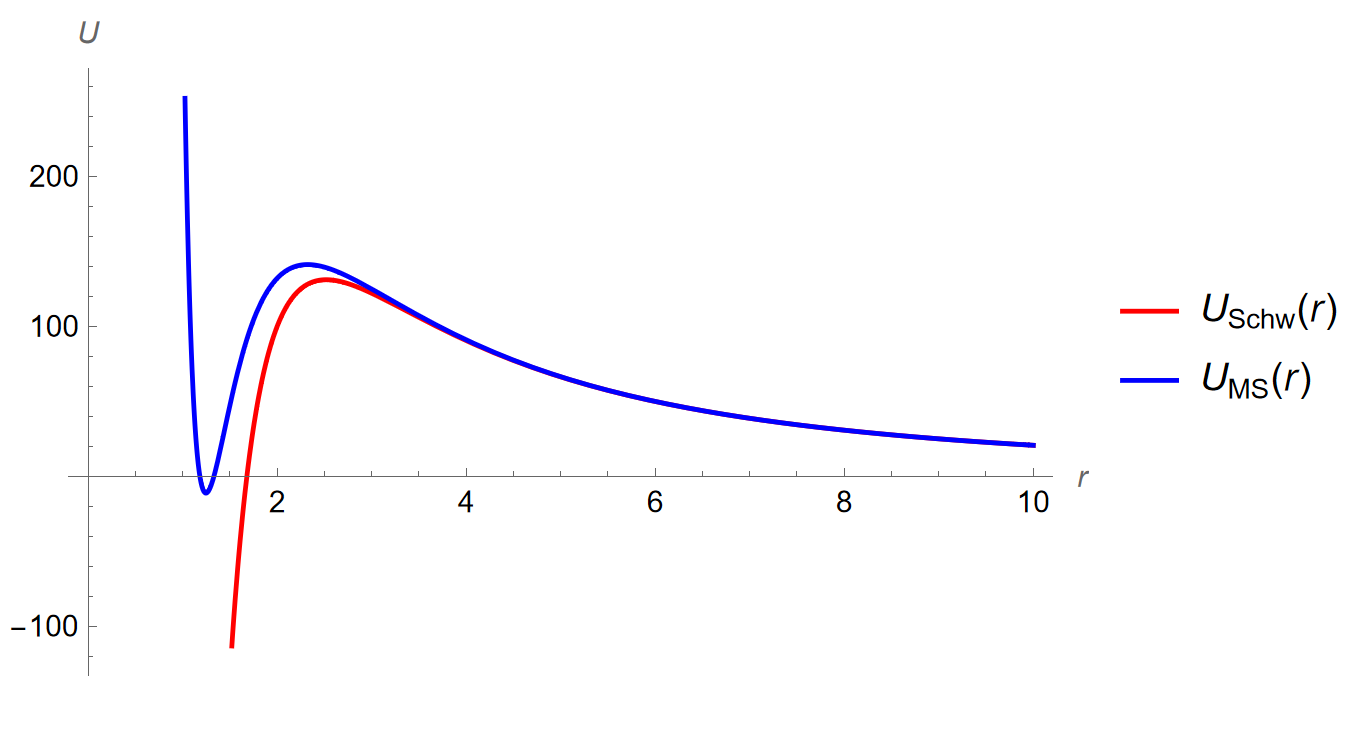}
\caption{The red curve $U_{\textrm{Schw}}$ is the effective potential of a SBH, and the blue one $U_{\textrm{MS}}$ is the effective potential of the QBH.}
\label{fig1}
\end{figure}

 In order for comparison with SBH, we choose the mass parameter $ M = 0.84 $, which is a little larger than $ M_{\textrm{min}} $, in all numerical figures in this section. The effective potentials of both the QBH and SBH are shown in \eqref{fig1}. It is obviously that the radius of the light ring of the QBH is a little smaller than that of the SBH. Explicitly, the radius of the light ring of QBH in this case is $ r_{l} \approx 2.323 $ and  the radius of the light ring of SBH in this case is $ r_{\textrm{Schw}} = 3M \approx 2.52$.

\begin{figure}[tbph] \label{fig2}. 
    \centering
    \includegraphics[width=8.5cm]{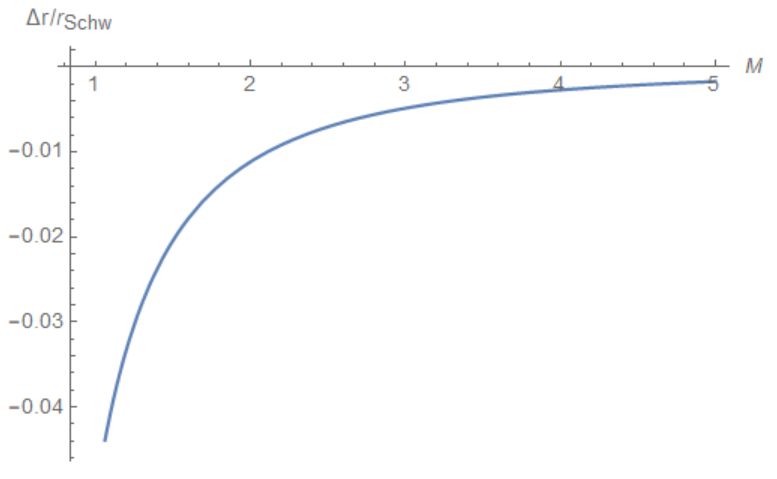}
    \caption{The deviation of light-ring radii of the QBH and SBH. $\Delta r=r_{l}-r_{\textrm{Schw}}$.}
\end{figure}

\begin{figure}[tbph]
    \centering
    \includegraphics[width=8cm]{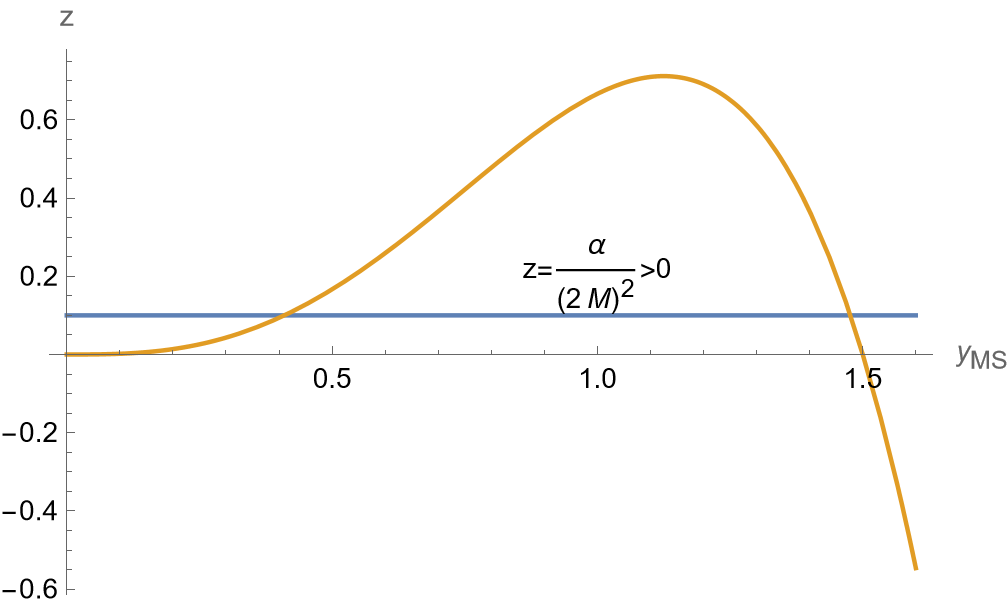}
    \caption{The function of $z$ and $y_{\textrm{MS}}$.}
    \label{fig5}
\end{figure}

The deviation of the radius of a QBH light ring from that of a SBH (with the same mass) decreases with the increase of the BH mass. This is shown in FIG.\eqref{fig2}. 
Although we have obtained the exact relation between the radius of the light ring of QBH and the quantum correction parameter $\alpha$ in Eq.\eqref{rLR},  it is easy to see in FIG.\eqref{fig2} that the quantum correction is much less than 1 even for a BH with Plank mass. Then we present a linear approximation for the light-ring radius with respect to the 
quantum correction parameter $\alpha$. Define two dimensionless parameter $y$ and $z$ as  
\begin{eqnarray}
    y = \frac{r}{2 M},~~ z = \frac{\alpha}{(2 M)^2}.
\end{eqnarray}\label{dimensionless}
For the dimensionless light ring radius $y_{\textrm{MS}}$, we have $y_{\textrm{MS}} =\frac{r_{l}}{2M}$. $y_{\textrm{MS}}$ and $z$ satisfy the following dimensionless equation
\begin{equation}
         z = - \frac{2}{3}\left(2 y_{\textrm{MS}}^4 - 3 y_{\textrm{MS}}^3\right).
\end{equation}
We plot this relation in FIG.\eqref{fig5}. When $ z= 0 (\alpha=0)$ , we have a nonzero solution $y_{\textrm{MS}} = \frac{3}{2} (r_{l} =3M)$, and this is just the case of a SBH. When $z > 0$, there are two positive real roots and the larger one is the light-ring radius. 
Since the quantum correction is much less than 1, we approximate $y_{\textrm{MS}}$ around $\frac{3}{2}$, and obtain 
\begin{equation}\label{linear}
        z \simeq - \frac{9}{2} \left(y_{\textrm{MS}} - \frac{3}{2}\right), 
\end{equation}
i.e.
\begin{equation}
r_{l} \simeq 3 M - \frac{\alpha}{9 M}.
\end{equation}

Next, we consider the critical impact parameter $b_{c}$ with which a light ray from spatial infinity approaches to the light ring of the QBH and has an arbitrarily large deflection angle.  
This critical impact parameter is determined by the following equation
\begin{equation}\label{eq14}
    \left(\frac{1}{b_{c}}\right)^{2} - \frac{1}{r_{l}^{2}} \left(1 - \frac{2 M}{r_{l}} + \frac{\alpha M^{2}}{r_{l}^{4}}\right) = 0,
\end{equation}
 and we have
\begin{equation}\label{shadow}
        b_c = \frac{r_{l}}{\sqrt{\left(1 - \frac{2 M}{r_{l}} + \frac{\alpha M^{2}}{r_{l}^{4}}\right)}}
\end{equation}
Taking a linear approximation with respect to the quantum correction parameter $\alpha$, we can obtain
\begin{equation}\label{linear_shadow}
        b_c = 3 \sqrt{3} M - \frac{\alpha}{6 \left(\sqrt{3} M\right)} + \mathcal{O}(\alpha^2).
\end{equation}
The first term is just the critical impact parameter of a SBH with mass $M$.

\section{Fine structure of photon rings}\label{section4}
\begin{figure}[tbph]
    \centering
    \includegraphics[width=9.5cm]{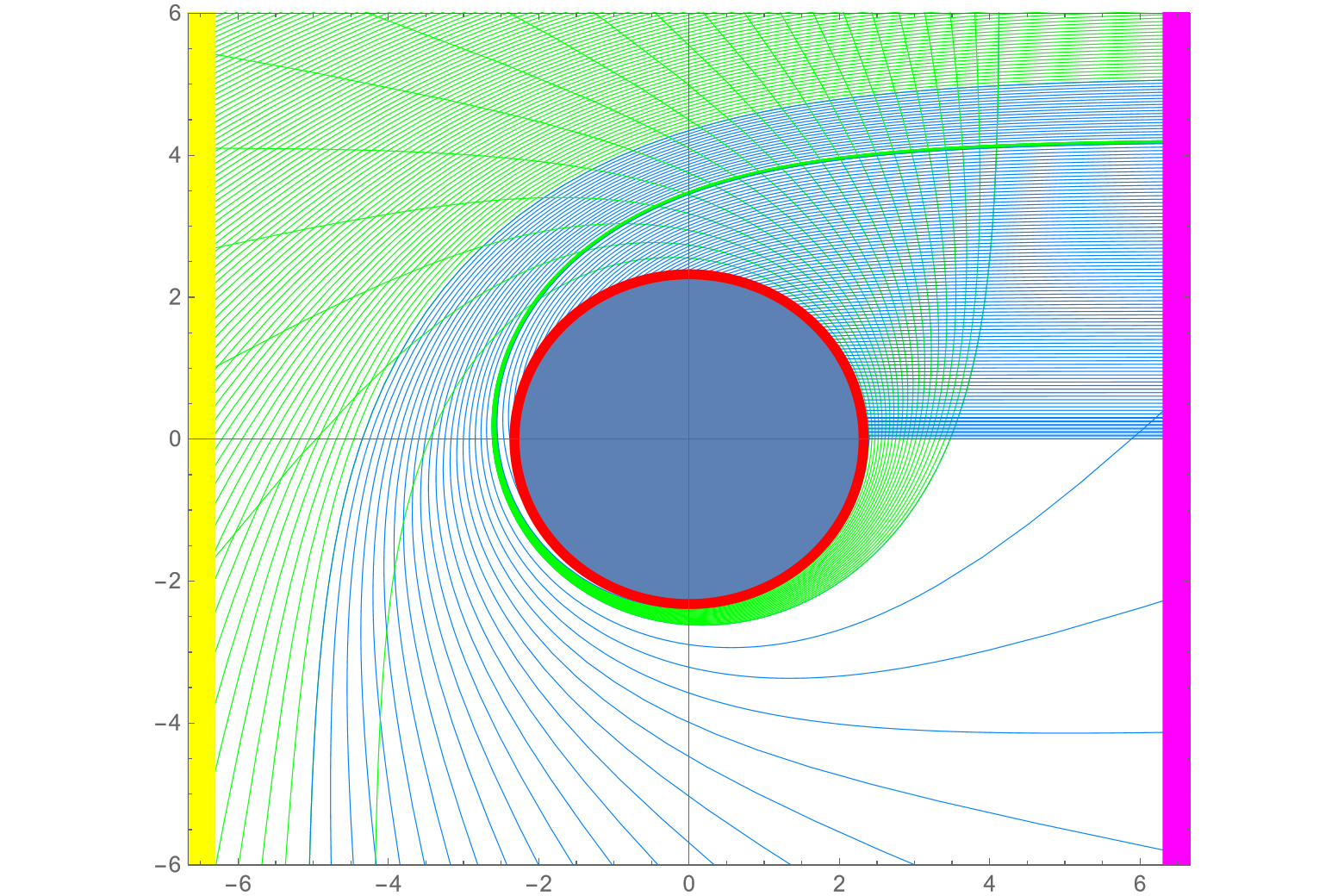}
    \caption{The shadow and photon ring. The red circle stands for the light ring, the yellow region on the left stands for the background light source and the purple region on the right stands for screen of the observer. The green lines indicate trajectories of light that can be observed on the screen and the blue lines indicate trajectories of light that can't be observed.}
    \label{Backgroundlight}
\end{figure}

In this section, we use the light-tracing method to derive the radius of the shadow of QBH in our model and discuss the bright photon ring structure in the shadow. The light rays with different impact parameters have different deflection angles. Here the deflection angle is defined as the total change in orbital plane azimuthal angle of a null geodesic. 
The deflection angle of a ``straight-line motion'' is $\pi$.  When the impact parameter $b$ is large enough, the deflection angle is a little larger than $\pi$, and the light trajectory intersects with the background light source. This kind of light form a bright region in the view of a distant observer. See the green lines outside the blue-line region in Fig.\eqref{Backgroundlight}.
 When the impact parameters become smaller and the values of the deflection angles are between $\frac{3}{2} \pi$ and $\frac{5}{2} \pi$, the trajectories of the light doesn't intersect with the background light source, and the relevant impact parameters correspond the first shadow region on the screen of a distant observer. See the blue lines in Fig.\eqref{Backgroundlight}.  
 When the values of the deflection angles are between $\frac{5}{2} \pi$ and $\frac{7}{2} \pi$, the trajectories of the light intersect with the background light source, and the relevant impact parameters correspond a bright ring region inside the shadow on the screen of a distant observer. See the blue lines in Fig.\eqref{Backgroundlight}. 
 
\begin{figure}[tbph]
    \centering
    \subfigure[$\varphi_{\textrm{def}} = \frac{3}{2}\pi$] {
        \label{Fig.sub.1}
        \includegraphics[width=3.2cm,height = 2.8cm]{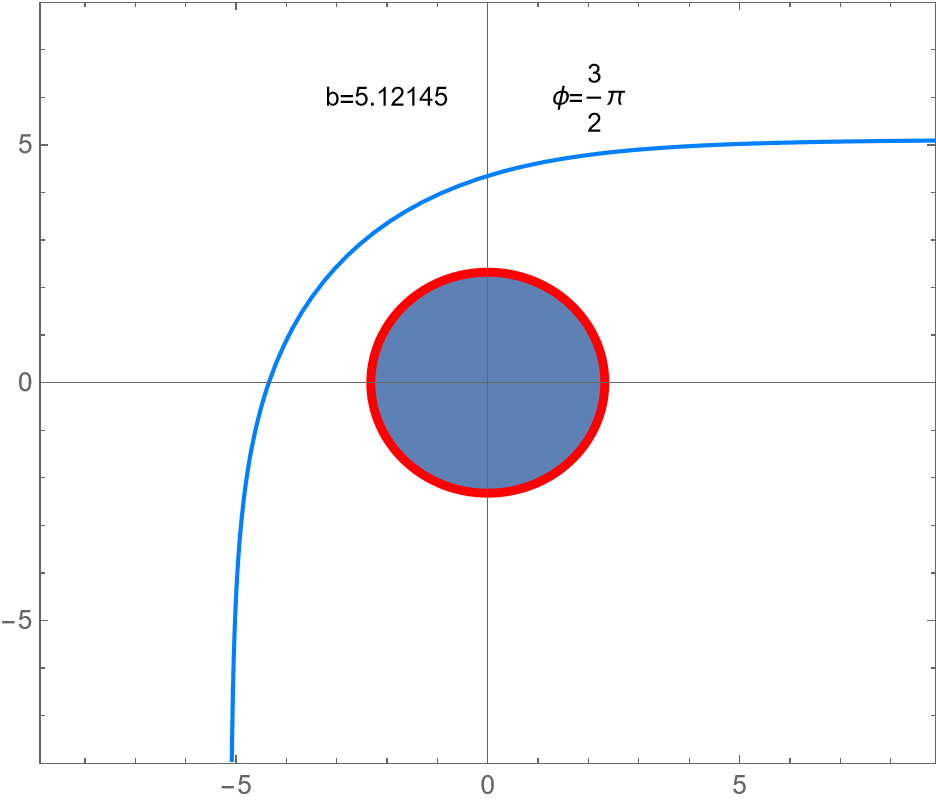}
    }
    \subfigure[$\varphi_{\textrm{def}} = \frac{5}{2}\pi$] {
        \label{Fig.sub.2}
        \includegraphics[width=3.2cm,height = 2.8cm]{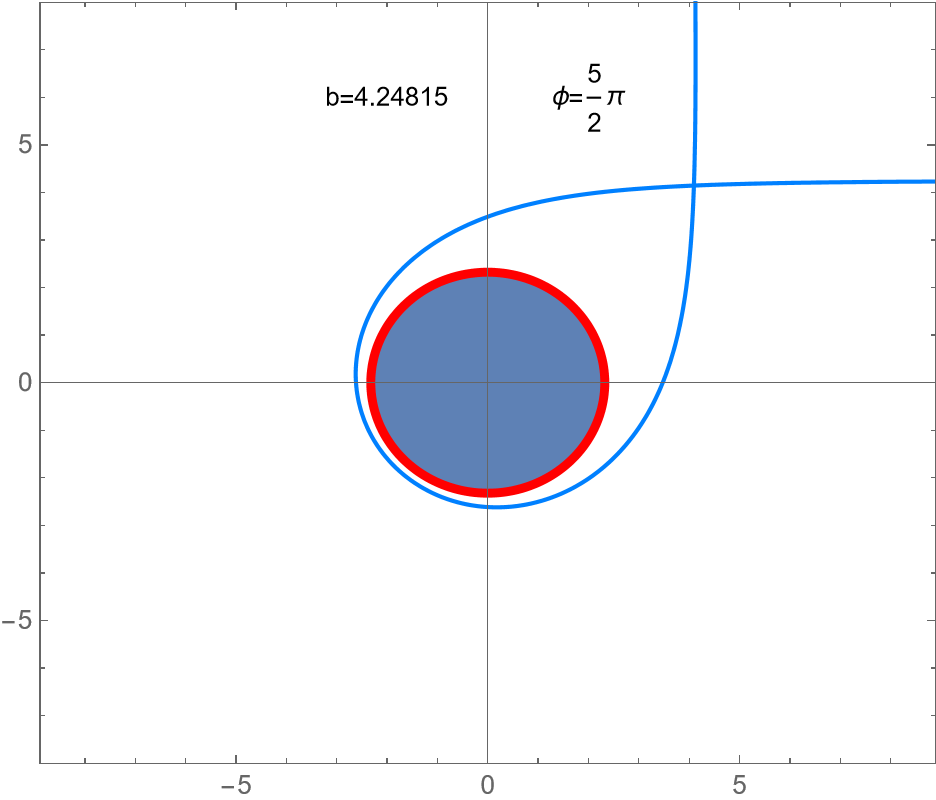}
    }
    \subfigure[$\varphi_{\textrm{def}} = \frac{7}{2}\pi$] {
        \label{Fig.sub.3}
        \includegraphics[width=3.2cm,height = 2.8cm]{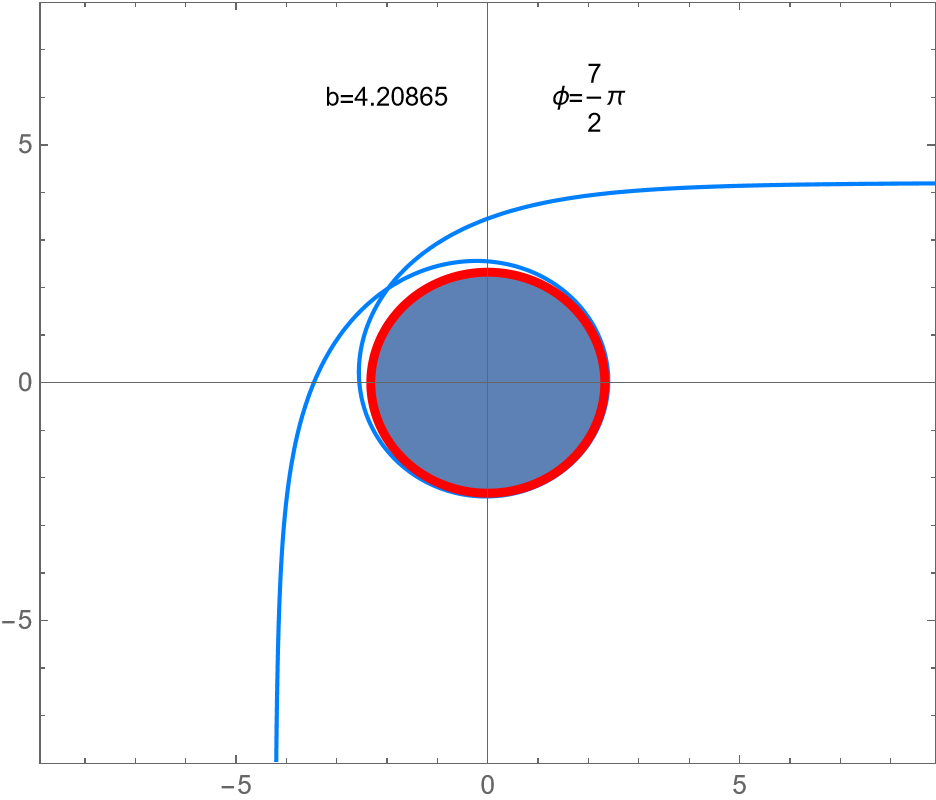}
    }
    \subfigure[$\varphi_{\textrm{def}} = \frac{9}{2}\pi$] {
        \label{Fig.sub.9}
        \includegraphics[width=3.2cm,height = 2.8cm]{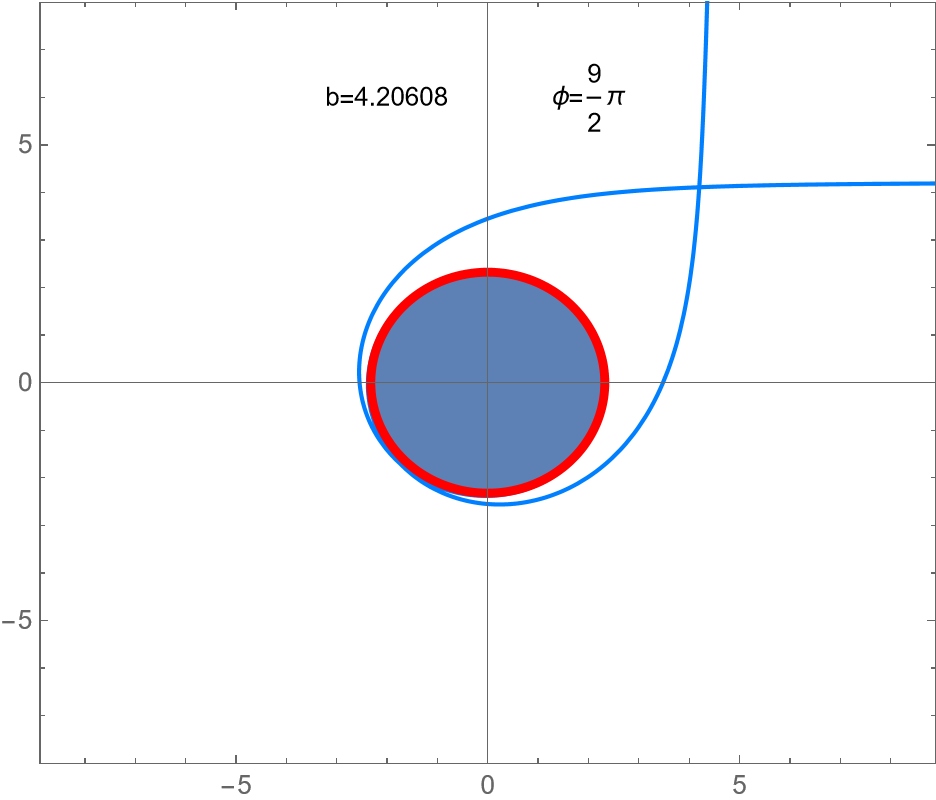}
    }
    \caption{Typical trajectories with different deflection angles.}
    \label{track}
\end{figure}

\begin{figure}[tbph]
    \centering
    \subfigure[The shadow and photon ring of the QBH.]{
        \label{MS_img}
        \includegraphics[width=3.5cm,height=3.5cm]{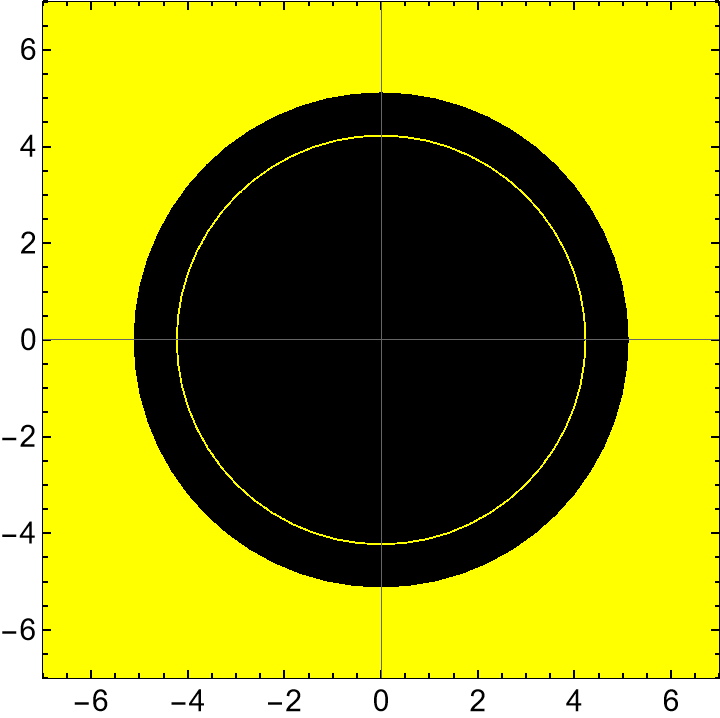}
    }
    \subfigure[The shadow and photon ring of the SBH.]{
        \label{Schw_img}
        \includegraphics[width=3.5cm,height=3.5cm]{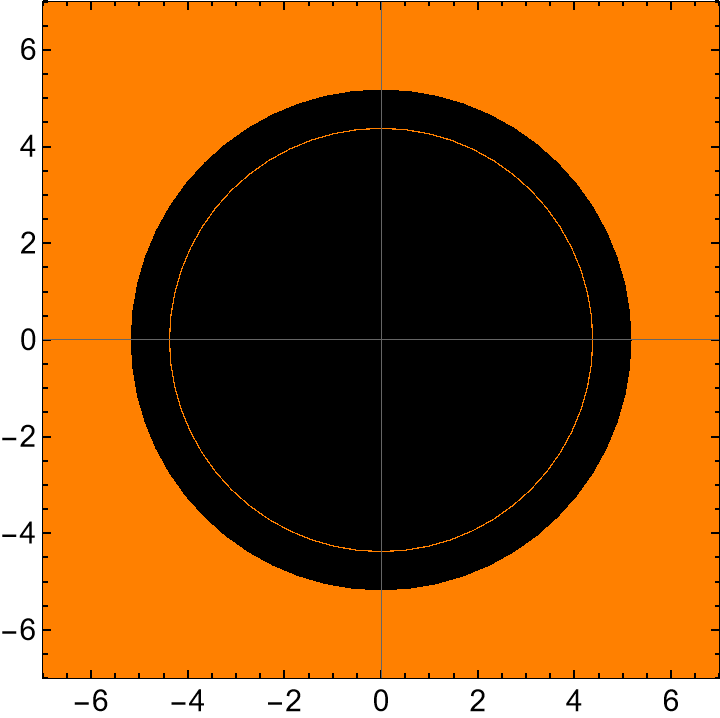}
    } 
    \subfigure[The comparison of the shadows and photon rings between the QBH and SBH.]{
        \label{img_compare}
        \includegraphics[width=4cm,height=4cm]{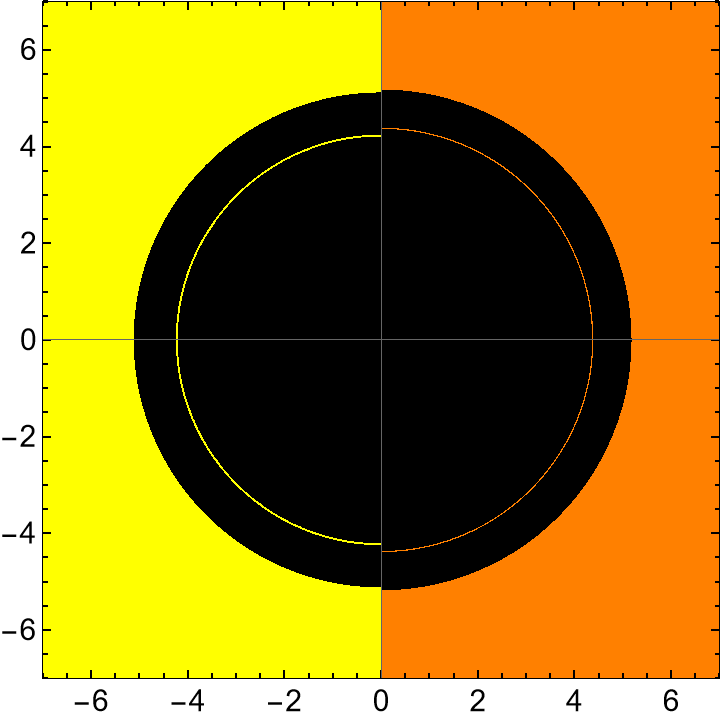}
    } \\
    \subfigure[A zoomed-in image from Fig.\eqref{MS_img} shows the fine structures of the photon ring of the QBH. One can see the $0^{\textrm{th}}$ order photon ring is much wider than the $1^{\textrm{st}}$ order photon ring.]{
        \label{img_fineStructureMS}
        \includegraphics[width=8cm,height=5cm]{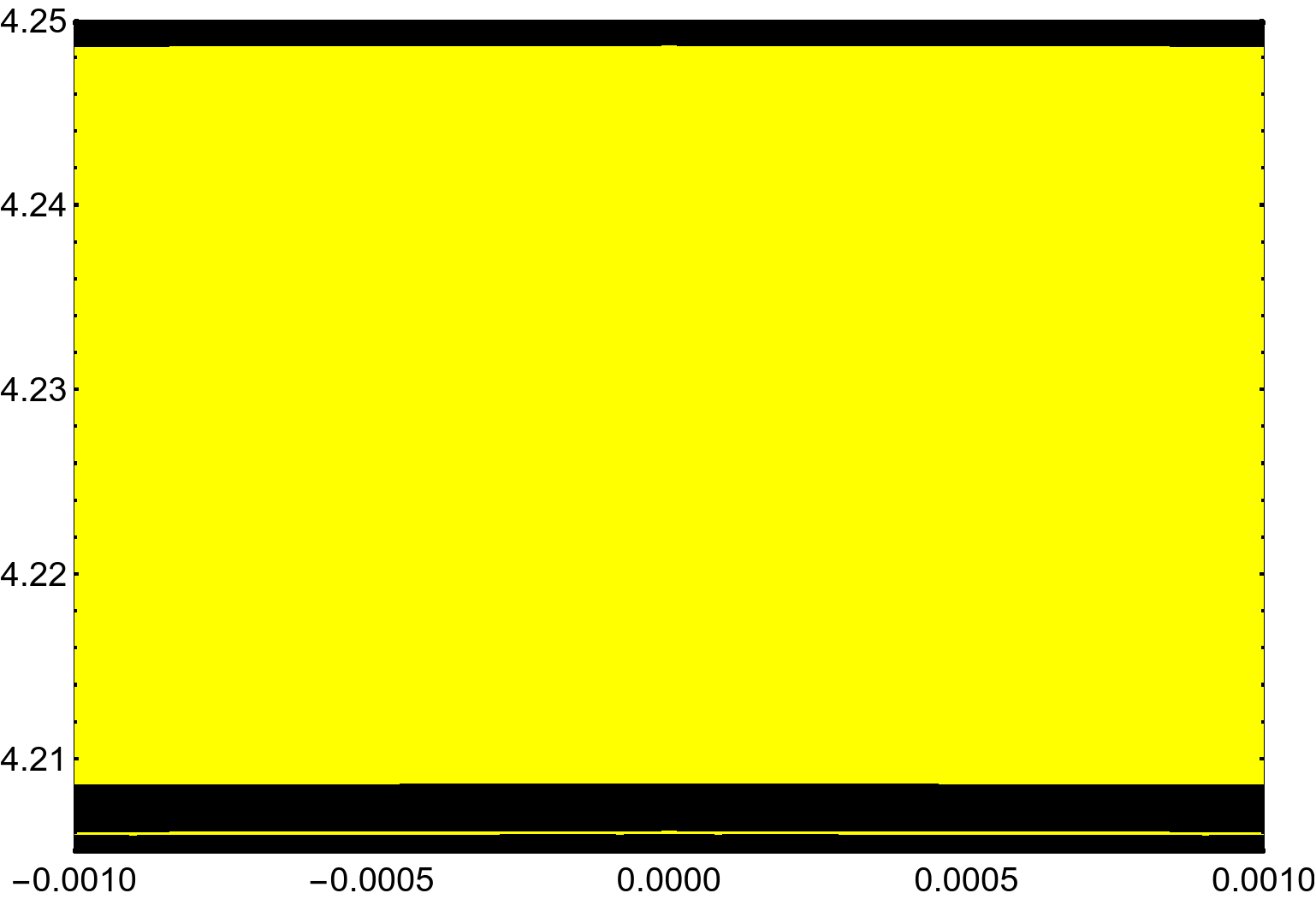}
    }
    \caption{The fine structure of the photon rings.}
    \label{img}
\end{figure}

As we mentioned before, when the impact parameter $b$ approaches to $b_c$, the deflection angles can be arbitrarily large. For $b<b_c$, the trajectories of the light fall into the inside region of the light ring and don't intersect with the background light source. These correspond to a shadow region on the screen of the distant observer.
For $b>b_c$, the trajectories with deflection angles belonging to the intervals $(\frac{3}{2}\pi + 2 k \pi, \frac{5}{2} \pi + 2 k \pi),k=0,1,2,... $ correspond to a $k^{\textrm{th}}$ order shadow region on the screen of the distant observer, and the trajectories with deflection angles belonging to the intervals $(\frac{5}{2}\pi + 2 k \pi, \frac{7}{2} \pi + 2 k \pi),k=0,1,2,... $ correspond to a $k^{\textrm{th}}$ order bright ring on the screen of the distant observer. In fact, except the $0^{\textrm{th}}$ order bright ring, the width of all higher order bright rings is tiny. This can be seen in the following Table.\ref{angleImpParaTable} and in Fig.\eqref{img_fineStructureMS}. The trajectories with typical deflection angles, $\varphi_{\textrm{def}}=\frac{3}{2}\pi,\frac{5}{2}\pi,\frac{7}{2}\pi,\frac{9}{2}\pi$, are shown in Fig.\ref{track}. 

\begin{center}
    \begin{table}[!h]
        \begin{tabular}{ | c | c | c | c | c | }
            \hline
            $\varphi_{\textrm{def}}$ & $3/2 \ \pi$ & $5/2 \ \pi$ & $7/2 \ \pi$ & $9/2 \ \pi$ \\
            \hline
            b & 5.121380595 & 4.248622455 & 4.208645804 & 4.205903279 \\
            \hline
            $\varphi_{\textrm{def}}$ & $11/2 \ \pi$ & $13/2 \ \pi$ & $15/2 \ \pi$ & $17/2 \ \pi$ \\
            \hline
            b & 4.205903279 & 4.205891983 & 4.205891240 & 4.205891190 \\
            \hline
        \end{tabular}
        \caption{\label{angleImpParaTable}Values of $b$ for some specific deflection angles.}
    \end{table}
\end{center}

The shadows on the screen of a distant observer of the QBH case and the SBH case are shown in Fig.\eqref{MS_img} and Fig.\eqref{Schw_img} respectively, and the comparison between them is shown in Fig.\eqref{img_compare}. One can see the radii of the shadow and the bright ring of the QBH are both a little smaller than that of the SBH. A demonstration of the $1^{\textrm{st}}$ and $2^{\textrm{nd}}$ order bright rings are in Fig.\eqref{img_fineStructureMS}.

\begin{figure}[tbph]
    \centering
    \includegraphics[width=7.5cm]{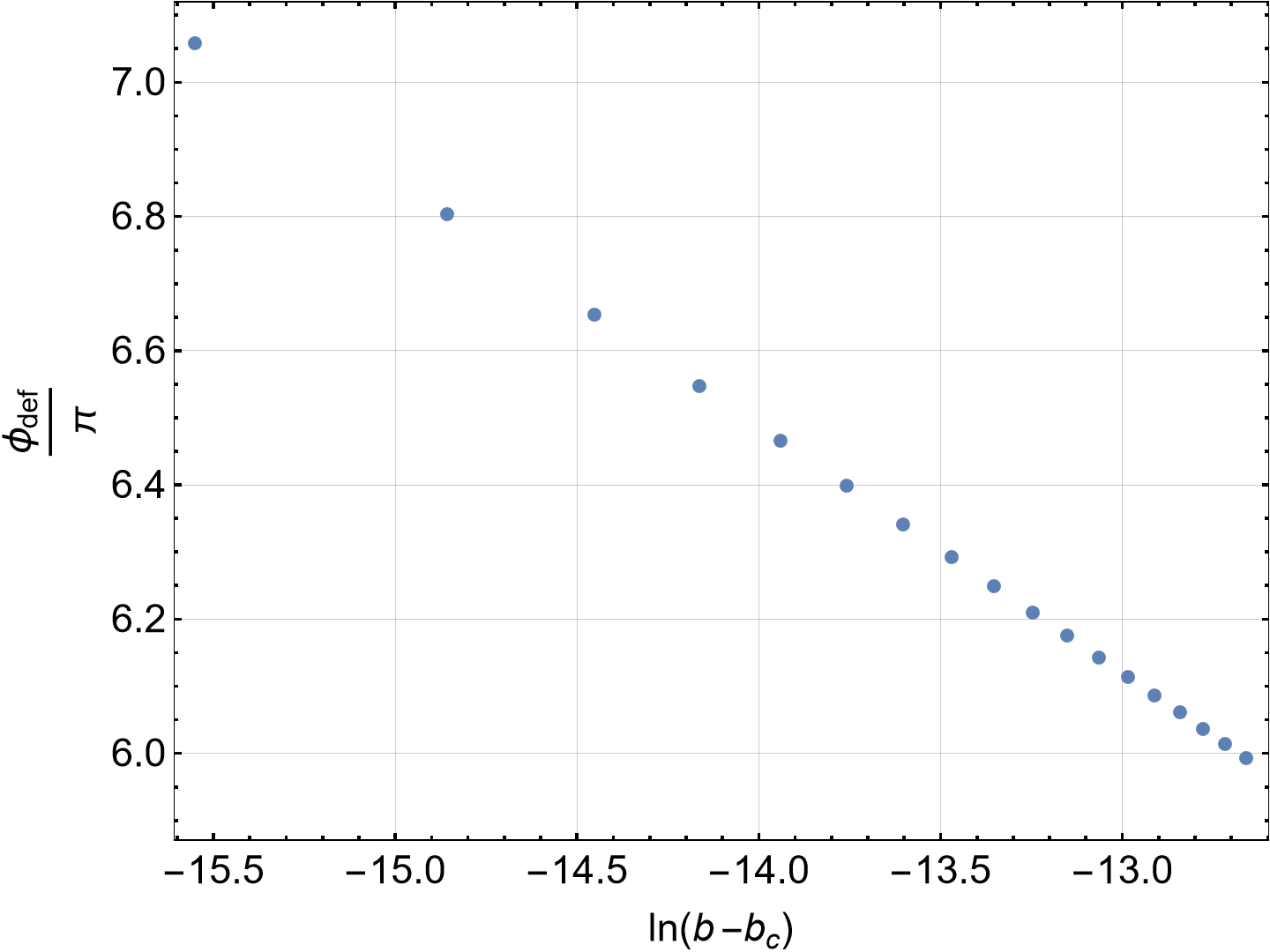}
    \caption{The numerical plot of $\varphi_{\textrm{def}}$  with respect to $\log(b - b_{\textrm{c}})$ when $b$ tends to $b_{\textrm{c}}$.}
    \label{the fine structure}
\end{figure}

Now let's consider the relation between the deflection angle $ \varphi_{\textrm{def}}$ and impact parameter $b$ when $b$ approaches to $b_c$.   From Eq.(\ref{governingEquationOfPhoton}), we can express the deflection angle of the light as a function of $b$,
\begin{equation}\label{4.5}
        \varphi_{\textrm{def}} = 2 \int_{r_{\textrm{turn}}}^{\infty}\frac{dr}{r^2\sqrt{\frac{1}{b^2} - U_{\textrm{MS}}(r)}},
\end{equation}
where $r_{\textrm{turn}}$ is the shortest distance from a trajectory to the BH. $r_{\textrm{turn}}$ is determined by the following equation
\begin{equation}
\frac{1}{b^2} = U_{\textrm{MS}}(r)|_{r=r_{\textrm{turn}}}.
\end{equation}
When $b=b_c$, $r_{\textrm{turn}}=r_{\textrm{MS}}$.
 Then we present an approximation of the above relation when $b\rightarrow b_{c}^+$. We introduce a cut-off $\delta$ ($\delta\ll r_{\textrm{turn}} $) to divide the above integral into two parts:
\begin{eqnarray}\label{appro}
        \int_{r_{\textrm{turn}}}^{\infty}\frac{dr}{r^2\sqrt{\frac{1}{b^2} - U_{\textrm{MS}}(r)}} = \int_{r_{\textrm{turn}}}^{r_{\textrm{turn}}+\delta}{\frac{dr}{r^2\sqrt{\frac{1}{b^2} - U_{\textrm{MS}}(r)}}} 
                 + \int_{r_{\textrm{turn}}+\delta}^{\infty}{\frac{dr}{r^2\sqrt{\frac{1}{b^2} - U_{\textrm{MS}}(r)}}}.
\end{eqnarray}
The second part gives a finite function $C_\delta(b)$ of $b$  even when $b = b_{\textrm{c}}$. When $b\rightarrow b_{c}^+$, the first part gives the divergent contribution, so we just need to consider the  approximation of the first part. When $b\rightarrow b_{c}^+$, we have $r_{\textrm{turn}} \rightarrow r_{l}$, and we can approximate $U_{\textrm{MS}}(r)$ as
\begin{equation}\label{4.7}
        U_{\textrm{MS}}(r) \simeq \frac{1}{b_{\textrm{c}}^2} - \frac{\omega^2}{2} {(r-r_{l})}^2,
\end{equation}
where $\omega^2=-U^{''}_{\textrm{MS}}(r_{l})$. Define a parameter $\epsilon = 1 - \frac{b_{c}^2}{b^2}(b>b_c)$ to measure the closeness to $b_{c}$ of $b$. 
Then the turning point $r_{\textrm{turn}}$ is determined by the following equation
\begin{equation}\label{4.8}
        \frac{\omega^2}{2}{(r_{\textrm{turn}} - r_{l})}^2 = \frac{\epsilon}{b_\textrm{c}^2}.
\end{equation} 
Considering $r_{\textrm{turn}} > r_{l}$,  we have
\begin{equation}\label{4.9}
        r_{\textrm{turn}} = r_{l} + \frac{\sqrt{2 \epsilon}}{\omega b_{\textrm{c}}}.
\end{equation}
 The first part of Eq.\eqref{appro} can be estimated as
\begin{eqnarray}\label{4.10}
        \int_{r_{\textrm{turn}}}^{r_{\textrm{turn}}+\delta}\frac{dr}{r^2\sqrt{\frac{1}{b^2} - U_{\textrm{MS}}(r)}} \simeq \frac{1}{r_{l}^2}\int_{r_{\textrm{turn}}}^{r_{\textrm{turn}}+\delta}\frac{dr}{\sqrt{\frac{1}{b^2} - U_{\textrm{MS}}(r)}} 
           \simeq - \frac{1}{\sqrt{2} \omega r_{l}^2} \zeta,
\end{eqnarray}
where
\begin{equation}\label{4.10.1}
    \zeta=\ln{\frac{\left(b_{\textrm{c}} \omega \delta-\sqrt{b_{\textrm{c}} \omega \delta (b_{\textrm{c}} \omega \delta + 2 \sqrt{2 \epsilon})} + \sqrt{2 \epsilon}\right)^2}{2 \epsilon}}
\end{equation}
Expand $\zeta$ with respect to $\epsilon$, we could obtain an approximation of Eq.\eqref{4.10}
\begin{equation}\label{4.11}
        \int_{r_{\textrm{turn}}}^{\delta}\frac{dr}{r^2 \sqrt{\frac{1}{b^2} - U_{\textrm{MS}}(r)}} \simeq - \frac{1}{\sqrt{2} \omega r_{l}^2} \log{(\epsilon)}.
\end{equation}
 By substituting $\epsilon = 1 - \frac{b_{\textrm{c}}^2}{b^2}$ into the above equation and considering the splitting in Eq.\eqref{appro}, we have
\begin{equation}\label{4.12}
        \varphi_{\textrm{def}} = - \frac{\sqrt{2}}{\omega r_{l}^2}\log{\left(b - b_{\textrm{c}}\right) + \widetilde{C}_\delta(b)}.
\end{equation}
Thus, we obtain a log-divergence for $\varphi_{\textrm{def}}$ when $b\rightarrow b_{c}^+$, which is similar to the SBH case \cite{Tsukamoto:2016jzh,Tsukamoto:2022uoz}. The function $\widetilde{C}_\delta(b)$ is the remaining finite contribution even when  $b=b_{c}$. In Fig.\eqref{the fine structure}, we plot $\dfrac{\varphi_{\textrm{def}}}{\pi}$ with respect to $\log(b - b_{\textrm{c}})$ based on the accurate integration Eq.\eqref{4.5}, and show that it is a linear curve when $b-b_{\textrm{c}} \to 0^{+}$, which is consistent with the approximation.

Finally, let's consider the coefficient $- \frac{\sqrt{2}}{\omega r_{l}^2}\equiv h$ of the log-divergence term in the above approximation \eqref{4.12}. According to the expression of the effective potential, Eq.\eqref{effectivePotentialOfPhoton}, and the definition of the dimensionless parameters, Eq.\eqref{dimensionless},  we can get
\begin{equation}\label{4.14}
      U_{\textrm{MS}}^{\prime\prime}(r)|_{r_{l}} = \frac{9 - 8 y_{\textrm{MS}}}{16 M^4 y_{\textrm{MS}}^5}.
\end{equation}
Since $U^{\prime\prime}(r_{l})=-\omega^2$, we have
\begin{equation}\label{omega}
        \omega = \sqrt{- \frac{9 - 8 y_{\textrm{MS}}}{16 M^4 y_{\textrm{MS}}^5}}.
\end{equation}
Substitute the linear approximation of $z$ in Eq.\eqref{linear} into the above equation, we obtain
\begin{equation}\label{4.15}
        \omega =  81 \sqrt{\frac{- 54 + 32 z}{M^4 (- 27 + 4 z)^5}}.
\end{equation}
Together with $r_{l} = 2 M y_{\textrm{MS}}$, we expand the coefficient $h=- \frac{\sqrt{2}}{\omega r_{l}^2}$ with respect to $z$,
\begin{equation}\label{4.18}
        h = h \left(0\right) + z \left.\frac{dh}{dz}\right|_{z=0} + \mathcal{O}(z^2),
\end{equation}
and we obtain the explicit expression for the coefficient $h$ at first-order quantum correction
\begin{equation}\label{4.19}
        h = 1 + \frac{2}{9} z + \mathcal{O}(z^2)=1 + \frac{\alpha}{18 M^2} + \mathcal{O}(\alpha^2).
\end{equation}
When the quantum correction $\alpha = 0$, we have $\left.h\right|_{\alpha = 0} = 1$ which is the corresponding coefficient in Schwarzschild case that was studied in \cite{Gralla:2019xty}. 

\section{Shadows of QBHs under various illumination backgrounds} 

In this section, we consider the shadows and photon rings of QBHs under three illumination models, and compare them with that of the Schwarzschild BH cases.
 Model 1 and 2 are similar to the models considered in the first and third rows in Fig.(6) in reference \cite{Gralla:2019xty}. A relatively thick accretion disk is considered in Model 3.
 
Now we first consider the brightness of the image of a BH and the surrounding emitting matter. One should integrate the emissivity along the light ray that reaches the screen \cite{Gralla:2019xty}. To simplify the calculations, we ignore the redshift effects since the relative brightness of different regions of the image is more important.   
Quantitatively, the brightness of a point in the image with light ray $\mathcal{C}$ is proportional to  
$B_\mathcal{C}=\int_\mathcal{C} \rho dl$.
Since the axial symmetry of our models, the above integral can be reduced to an 2-dimensional integral. We choose $(r,\varphi)$ as the polar coordinates.   
With the equation of motion of photon,  the brightness corresponds to light a path with impact
parameter $b$ is proportional to  
\begin{eqnarray}
B(b)=\frac{1}{b}\int_{0}^{\varphi_{\text{def}}(b)}\rho(r,\varphi)r^2f^{-1/2}d\varphi,
\end{eqnarray}
where $r$ is a function of $\varphi$ determined by Eq.\eqref{governingEquationOfPhoton}.

In Model 1, the background light is emitted from an optically thin but geometrically thick disk region near the black hole. The emission disk extends from the BH horizon to  spatial infinity. The thickness of the disk is chosen as three times of the radius of the light ring radius, i.e. $3r_l$. We assume the disk emits
isotropically with uniform brightness. It is also considered that the screen is infinitely far away from the BH and is perpendicular to the rotationally symmetric axis of the disk.  

In Model 2, the background light is emitted from a spherical cloud near the black hole. The emission region extends from the photon sphere to the exterior region of the BH. The emission in this model is locally isotropic, but the emitted density is exponentially decaying with respect to the radial coordinate, i.e. the  emission intensity $\rho$ is chosen as $\rho = \rho_0\exp{(- r^{2}/M^2)}, r \geq  r_{l}$.

In Model 3, the background light is emitted from a relatively thick accretion disk. The emission profile is chosen as
$\rho =\rho_0 (r - r_{\text{Hz}})^{-3}, r \geq  r_{l}$, $r_{\text{Hz}}$ is the radius of the event horizon of the black hole. The thickness of the disk is comparable with the radius of the black hole horizon $r_{\text{Hz}}$.

 \begin{figure}[tbph]
		\centering
		\subfigure[Model 1 : Thick disk]{
			\label{Comparism-Model1}
			\includegraphics[height=3.3cm]{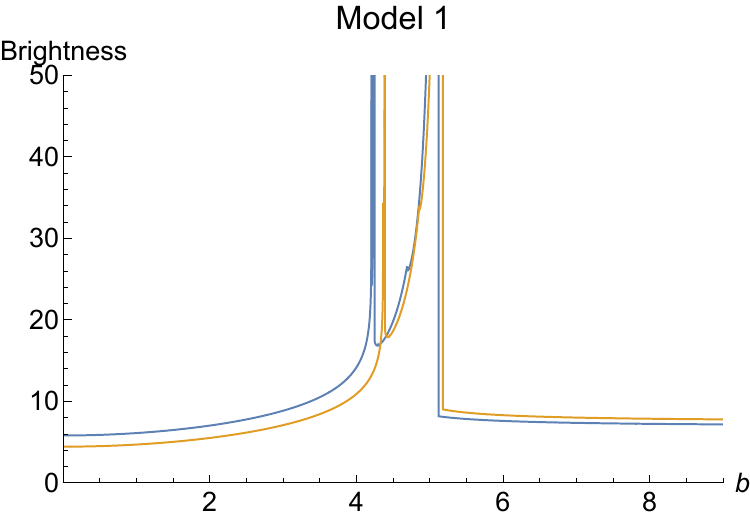}
		}
		\subfigure[Model 2 : Spherical cloud]{
			\label{Comparism-Model2}
			\includegraphics[height=3.3cm]{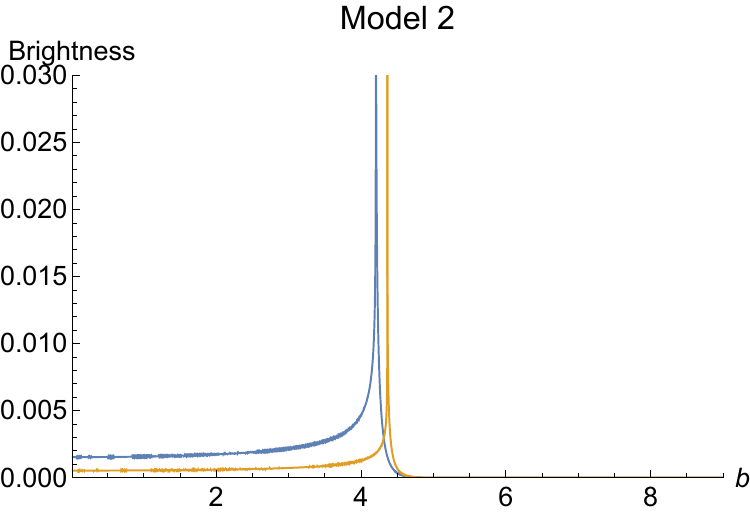}
		} 
		\subfigure[Model 3 : Accretion disk]{
			\label{Comparism-Model3}
			\includegraphics[height=3.3cm]{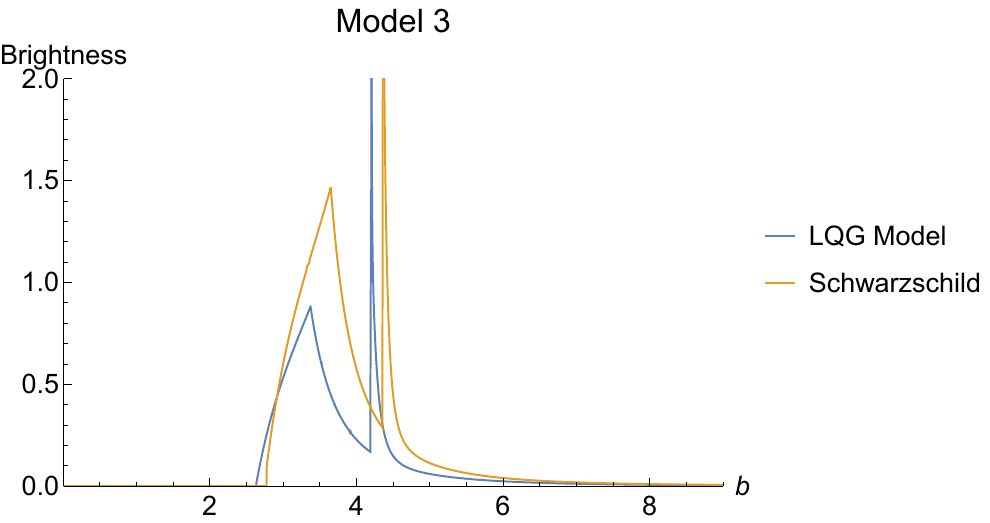}
		} 
		\subfigure[Comparism of images between the LQG (LHS) and Schwarzschild's (RHS) metric with model 1.]{
			\label{Comparism-1.1}
			\includegraphics[width=5.2cm]{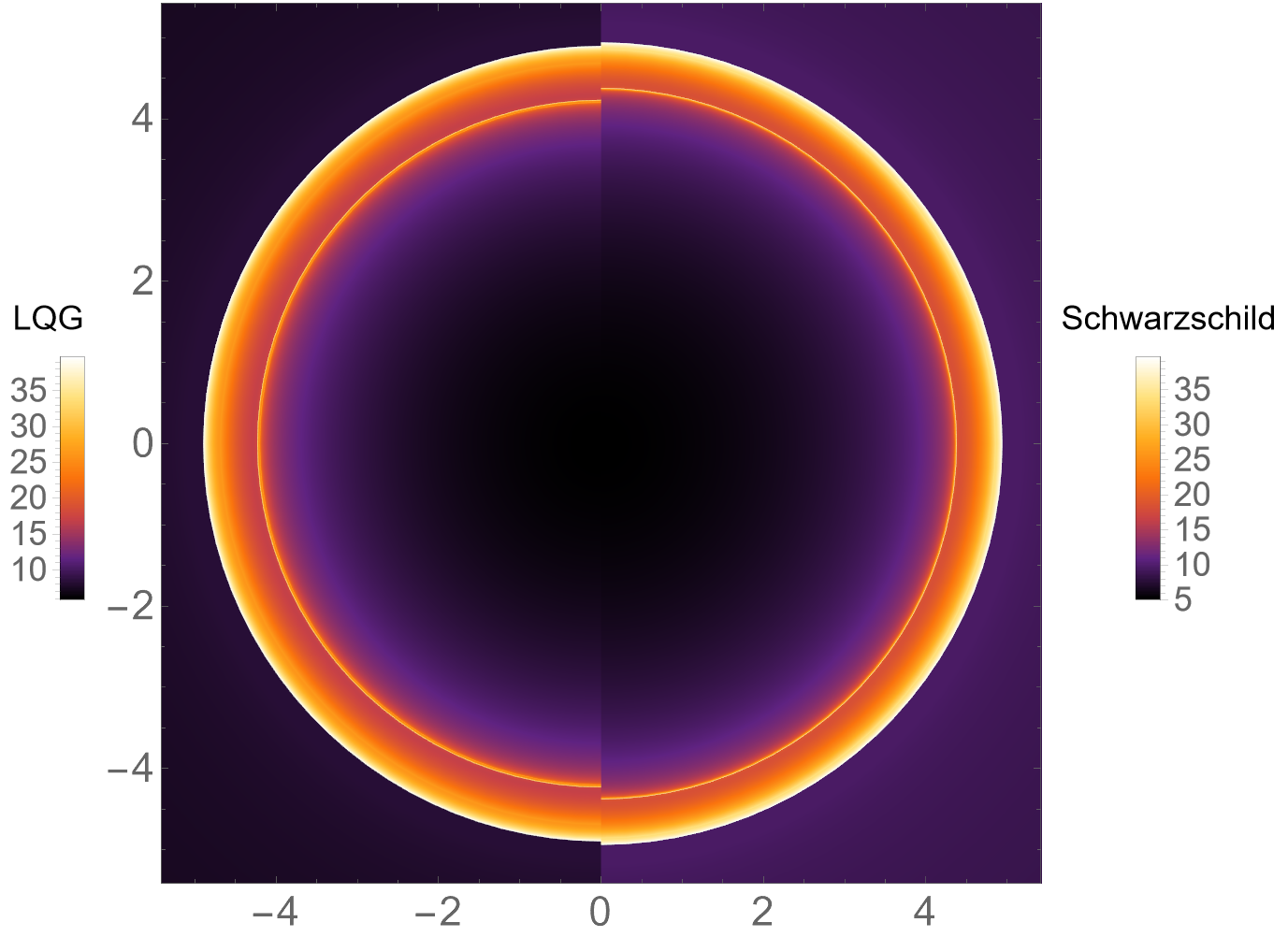}
		}
        \subfigure[Comparism of images between the LQG (LHS) and Schwarzschild's (RHS) metric with model 2.]{
			\label{Comparism-2.1.1}
			\includegraphics[width=5.2cm]{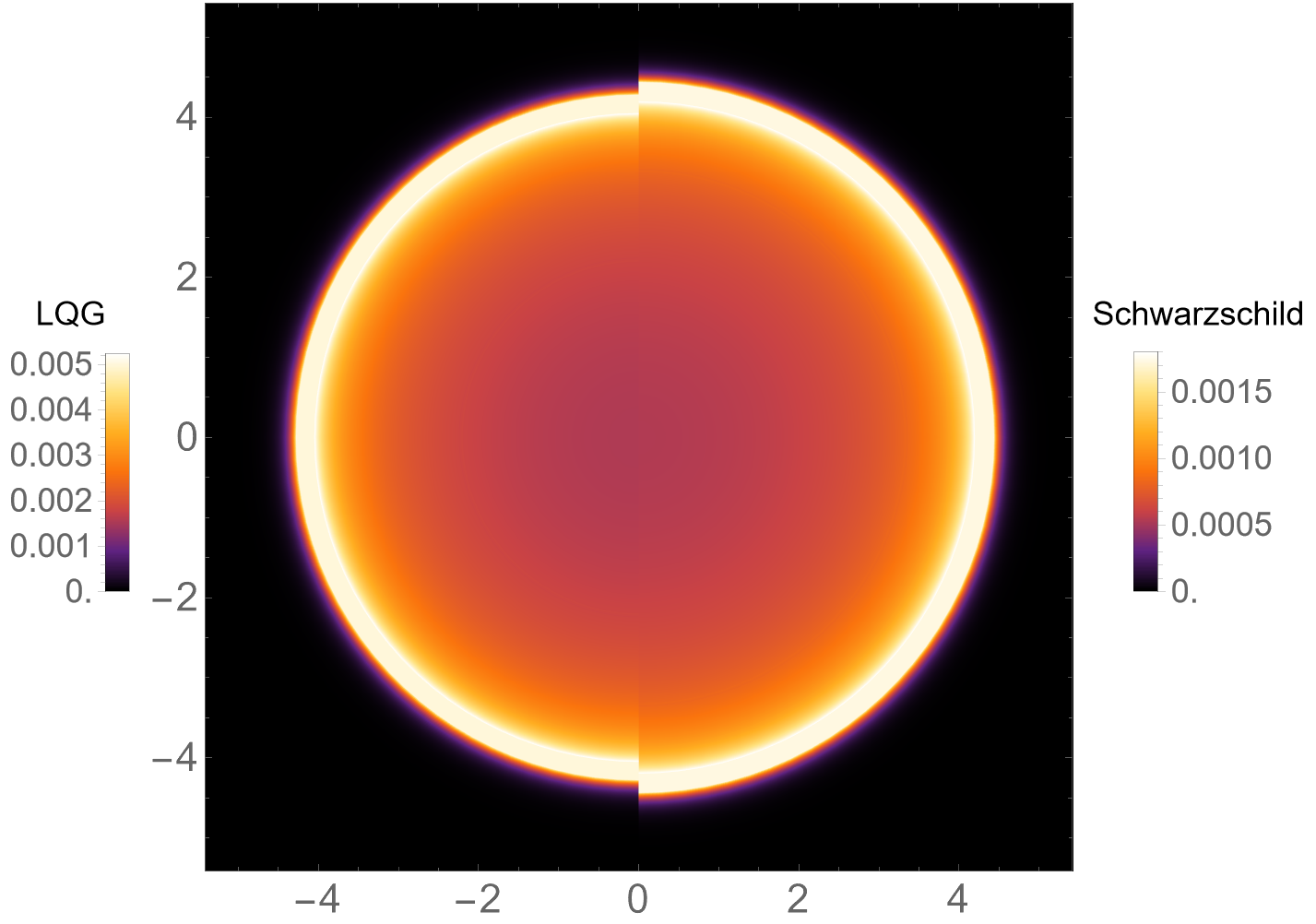}
		}
       \subfigure[Comparism of images between the LQG (LHS) and Schwarzschild's (RHS) metric with model 3.]{
			\label{Comparism-3.1}
			\includegraphics[width=5.2cm]{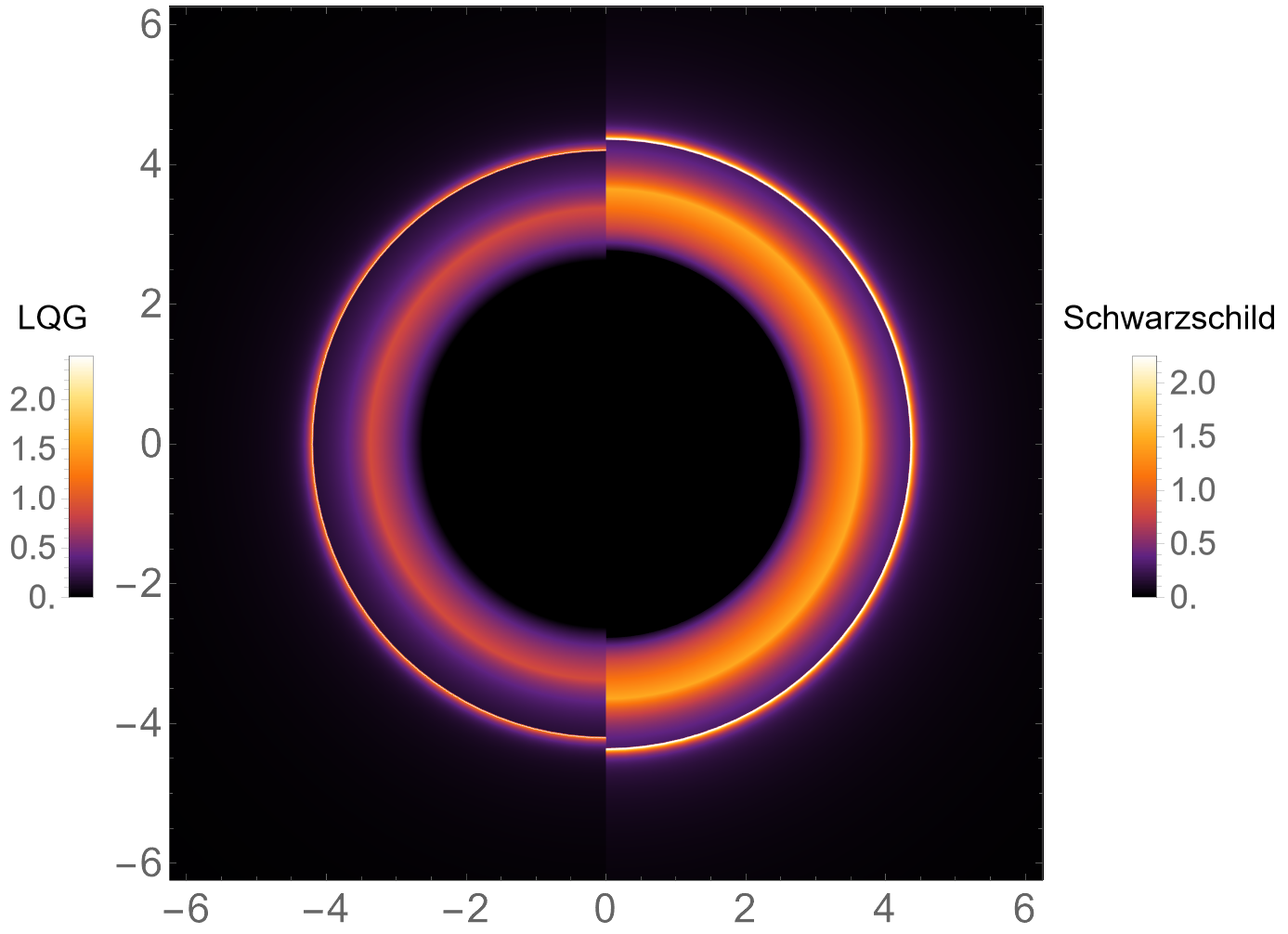}
		}
		\caption{Shadows and photon rings of the QBH from LQG model in three illumination models and comparisons with that of the SBH. In each of the lower three figures, the left-hand side (LHS) is the image for QBH and the right-hand side (RHS) is for SBH. }
		\label{Comparism-1}
	\end{figure}
\begin{figure}[tbph]
		\centering
	\label{Model3-changethick}
	\includegraphics{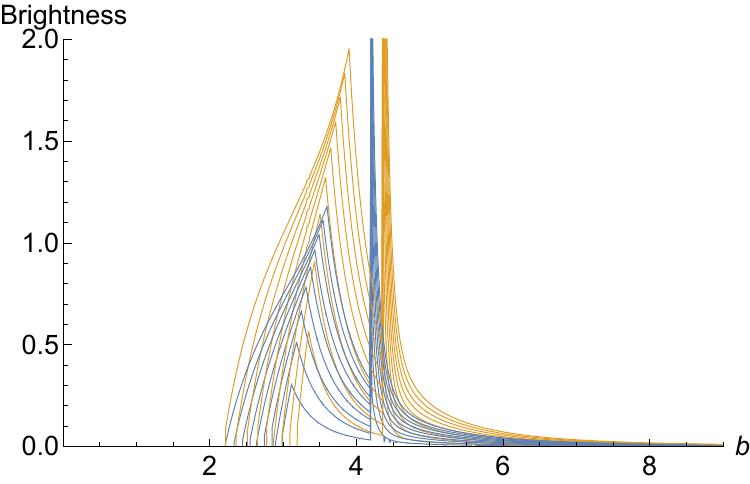}
	\caption{The blue curves are for the brightness of the QBH and the orange curves are for the SBH. In each kind of curves, from bottom to top, the thickness of accretion disk is respectively $ 0.2 r_{\textrm{Hz}}, 0.4 r_{\textrm{Hz}}, 0.6 r_{\textrm{Hz}}, 0.8 r_{\textrm{Hz}}, 1.0 r_{\textrm{Hz}}, 1.2 r_{\textrm{Hz}}, 1.4 r_{\textrm{Hz}}, 1.6 r_{\textrm{Hz}}, 1.8 r_{\textrm{Hz}}$.}
	\end{figure}

For figures in Fig.\eqref{Comparism-1}, we choose $\rho=1$ in Model 1 and $\rho_0=1$ in Model 2 and Model 3. The thickness of the accretion disk in the third column is chosen to be equal to the radius of event horizon $r_{\text{Hz}}$. The three plots in the first row show the variation of brightness with respect to the impact parameter $b$ (different paths) in the three illumination models. We also compare the brightness of the images of the QBH from LQG  and SBH in each illumination model. The blue line is for the QBH and the orange line for the SBH. 
In Model 1 and Model 2, the differences between the two BHs are both qualitatively and quantitatively small. 
In Model 3, there are two brightness peaks in the images of the two BHs. The outer peak is located near the critical impact parameter. The difference between the images of the two BHs is qualitatively small but quantitatively large for the inner peaks. This brightness peak of SBH is almost twice of that of the QBH. This is interesting since it means we could distinguish the QBH from the SBH under certain illumination conditions. In Fig.\eqref{Model3-changethick}, we vary the thickness of accretion disk from $0.2 r_{\textrm{Hz}}$ to $1.8 r_{\textrm{Hz}}$ with step $0.2 r_{\textrm{Hz}}$ and plot nine sets of curves, which show that the inner brightness peak of SBH is always almost twice of that of the QBH in the disklike accretion models. 

The second row shows the images of shadows and photon rings of the QBH in three illumination models and the comparisons with that of the SBH cases. 
The photon rings in the shadows are clear in Model 1 and Model 3  in the disklike emission cases. However, Model 2 has a spherical emission and the photon-ring feature is not obvious. 
 
\section{Conclusion}\label{section5}
We study the shadows and the photon ring structures of a quantum black hole arising from a loop quantum cosmology model. We calculate the radius of the light ring of the quantum black hole, and give an approximation of it, $r_{l} \simeq 3 M - \frac{\alpha}{9 M}$, at the first order of the quantum correction parameter $\alpha$. Considering a simple model where we assume the quantum black hole is backlit by a large and distant plane of uniform and isotropic emission, we calculate the radius of the shadow, which is $ r_s = 3 \sqrt{3} M - \frac{\alpha}{6\left(\sqrt{3} M\right)} $ at the first order of the quantum correction parameter. We then study the structure of the photon ring, and compare the shadow and photon ring of the quantum black hole with that of the SBH. The quantum correction term  makes the radius of the shadow smaller than that of the Schwarzschild black hole, which is consistent with the claim in \cite{Lu:2019zxb}. We also estimate the relationship between the deflection angle $\varphi_{\textrm{def}}$ and the impact parameter of a light ray. When $b$ is approaching to the critical value $b_c$, we obtain a logarithmic relationship $\varphi_{\textrm{def}} \simeq \frac{\sqrt{2}}{\omega r_{l}^2}\log\left(b - b_{\textrm{c}}\right)$. 

We then plot the images of the shadows and photon rings of the QBH under three illumination conditions. And we also compare the images of QBH with that of the SBH in each illumination models. In Model 1 and Model 2, there are only small quantitative difference between the images of the two kinds of BHs. 
In Model 3, it is found that there are two brightness peaks in the images of the two kinds of BHs. In each image of the BHs, the outer peak is located around the critical impact parameter. And, we also plot nine sets of brightness curves for QBH and SBH in Fig.\eqref{Model3-changethick} with respect to different thickness of the accretion disks. It is found that the inner brightness peak of the SBH is always almost twice of that of the QBH. This means we could distinguish the QBH and the SBH under some specific illumination conditions via the shadow images.

\section*{Acknowledgments}\label{Ack}
The authors thank Jinbo Yang, Zhan-Feng Mai and Cong Zhang for useful discussions and communications. This work is partially supported by Guangdong Major Project of Basic and Applied Basic Research (No.2020B0301030008).

\end{document}